\documentclass[11pt,a4paper]{article}
\usepackage[utf8]{inputenc}
\usepackage[english]{babel}
\usepackage{amsmath}
\usepackage[square,super,sort&compress,numbers]{natbib}
\usepackage{amsfonts}
\usepackage{amssymb}
\usepackage{graphicx}
\date{}
\usepackage{floatrow}
\graphicspath{ {Figures/}}

\title{Overcoming Distrust in Solid State Simulations: Adding Error Bars to Computational Data}
\author{Francesca Peccati,$^{*,a}$ Rub{\'e}n Laplaza$^{a,b}$ and Julia Contreras-Garc\'{i}a$^{*,a}$}

\begin{document}
\maketitle
\thispagestyle{empty}
\begin{center}
$^{a}$ Sorbonne Universit{\'e}, CNRS, Laboratoire de Chimie Th{\'e}orique, LCT, F 75005 Paris, France\par
\vspace{0.3cm}
$^{b}$ Departamento de Qu{\'i}mica F{\'i}sica, Universidad de Zaragoza, 50009 Zaragoza, Spain\par
\vspace{0.5cm}
* email: fpeccati@lct.jussieu.fr, contrera@lct.jussieu.fr
\end{center}


\subsection*{Abstract}
X-ray and neutron diffraction are well-established techniques for structural determination, whose success allowed the development of materials science. At the same time, simulation techniques are providing with each passing day a deeper insight into the structure and properties of materials. Two main obstacles appear for the cooperation of simulation and experiment. On the one hand, the frequent lack of a degree of uncertainty associated with calculated data. On the other, the concomitant underlying feeling that calculation parameters can be tuned with the explicit aim of matching the experimental results, even at the expense of the quality of the simulation. Without the definition of an error bar for estimating the precision of the calculation, direct comparison of calculated and experimental data can lack physical significance. In this contribution, we employ the well known delocalization error of DFT and HF to develop a simple and robust procedure to quickly estimate an error bar for calculated quantities in the field of solid state chemistry. First, we validate our model on one of the simplest properties of a solid, the geometry of its unit cell, which can be determined experimentally with high accuracy. In this case, our computational window is too large to provide a useful error bar. However, it provides computational material scientists with a pointer on how much a given system is affected by the method of choice, i.e. how much it is sensible to parameter tuning and how much care should be taken in doing it. Then, we move to another quantity which has a greater experimental uncertainty, namely transition pressure, and show that our approach can lead to error bars comparable to experiment. Hence, both experiment and theory can be compared on an even basis taking into account the uncertainty introduced by the scientist, both in the measuring conditions and the tuning of computational parameters.

\section{Introduction}
More than a century has passed since the first determination of the structure of a crystal (W.H. and W.L. Bragg, 1913).\cite{early_cryst} Since then,  X-ray and neutron diffraction have developped into maturity, providing a virtually unlimited access to high resolution structures, spanning not only over a wide range of material complexity but also over a wide range of temperature and pressure conditions. This technological advance was essential for the development of all branches of material science.  The evolution of material science has also profited from the parallel development of simulation techniques, which made accessible not only "bulk" features such as band structure, phase transitions mechanisms and defectivity, but also surface properties, including formation energy and reactivity, which are of paramount importance for the rationalization of heterogeneous catalysis.\cite{dov2005}\par 
At this point, a close collaboration between simulation and experiment has become routine in the mutual validation of data.
What often hinders this joint effort is the lack of a degree of uncertainty associated with simulated data, which are usually presented as naked numbers, without an associated precision, contrarily to experimental values, which are generally accompanied by a range that estimates the precision of the measurement. This often leads, in the comparison of experimental and calculated results,  to shady situations in which objectiveness succumbs to personal interpretation (and wishful thinking): is it acceptable to  assert a good agreement between theory and experiment when the calculated value falls just narrowly out of the experimental error range? The temptation to tamper with simulation parameters to have the calculated value nicely falling within the experimental range is hard to resist and gives rise to suspicion in the interaction between computational and experimental chemists.\cite{patrick_ipea_2017,bias_physics_2005}
In this work, we aim at changing the paradigm of the experiment-computational interaction, presenting in a clear way how much of the uncertainty on selected properties depends on the method employed in the calculations.\par  
To introduce the problem, we will start with one of simplest features of a crystal: its cell parameters.  Nowadays, a routine X-ray diffractometer measurement provides the cell parameters with an excellent precision, the uncertaintanty being as low as some parts in 10\textsuperscript{-4}, even for organic crystals, which can be lessened by a further order of magnitude by employing special techinques.\cite{precision_Xray} The case of NaCl is emblematic. In the first work by Bragg, dating back to 1913, the value of the cell parameter $a_{expt}$ was set to 4.45 \AA, only to be corrected in the same year to the value of 5.62 \AA, which is extremely close to the currently accepted value of 5.6401 \AA. \cite{early_cryst,Bragg13} Owing to its simplicity, the geometry of the unit cell is often one of the first quantities that are checked in the comparison between computation and experiment. In this situation, what is the result of the comparison between $a_{expt}$ = 5.6401 \AA~  and an hypothetical calculated value $a_{calc}$  = 5.7835 \AA~ with no error bar? With such a difference between $a_{calc}$ and $a_{expt}$, can we trust the computational technique employed to represent acceptably the real crystal, and therefore draw conclusions based on the physical insight provided by the simulation? To answer this question we have to investigate the sources of error that affect the value of the calculated cell parameter. A first consideration is that the geometry optimization of a solid, which yields the cell parameters, does not have an associated random error except the numerical one associated to arithmetic operations, which is negligible. This means that irrespective of the number of repetitions, a geometry optimization with the same starting point, method and simulation parameters will always converge to the same structure, and thus to the same $a_{calc}$  value. As a consequence, all the errors in this kind of calculation are systematic and therefore hard to eliminate. The main sources of systematic error are the level of theory employed (the combination of method and basis set) and other less evident variables, such as the sampling of the reciprocal lattice, integral truncation criteria and grids. We can group these errors into two groups, which we will call discretization and modelization errors. Discretization errors cover those errors coming from the finite treatement of infinite series: basis set, sampling and truncation.\cite{discretization}. 
Model errors refer to the method - the physical model used to describe a real system. Under this umbrella we have wavefunction and Density Functional Theory (DFT) methods. Whereas the first ones can build systematic improvements adding correlation on top of Hartree-Fock (HF), the latter are not systematic-improvement prone.
However, wavefunction correlated methods are not generally available for solid state, so that basically all material science computations are done within the DFT framework. Within this framework, many exchange-correlation functionals are available. However, the increase in computational cost and theoretical involvement of the functional does not ensure better results at all. This means that computational material scientists are left with the choice of functional and no security whatsoever of how the method is affecting the results (i.e. a more expensive functional will not necessarily lead to a better result). 
Having fixed all the remaining degrees (which can be systematically improved), we will now focus on the uncertainty related to working within the DFT framework, which is not predictable in advance, trying to answer the question: how is the model (functional) affecting the results?
For this, one first needs to understand the main errors coming from solid state computations. These have been summarized as: non-covalent interactions, strongly correlated systems and delocalization error.\cite{yang} In the absence of non-covalent interactions or strongly correlated systems (which are easy to identify), the main source error in DFT simulations is the ever-present delocalization error.\cite{Srebro2014} 

\begin{figure}[H]
\begin{minipage}[c][2.75cm][c]{0.5\textwidth}
\begin{center}
\includegraphics[width=\linewidth]{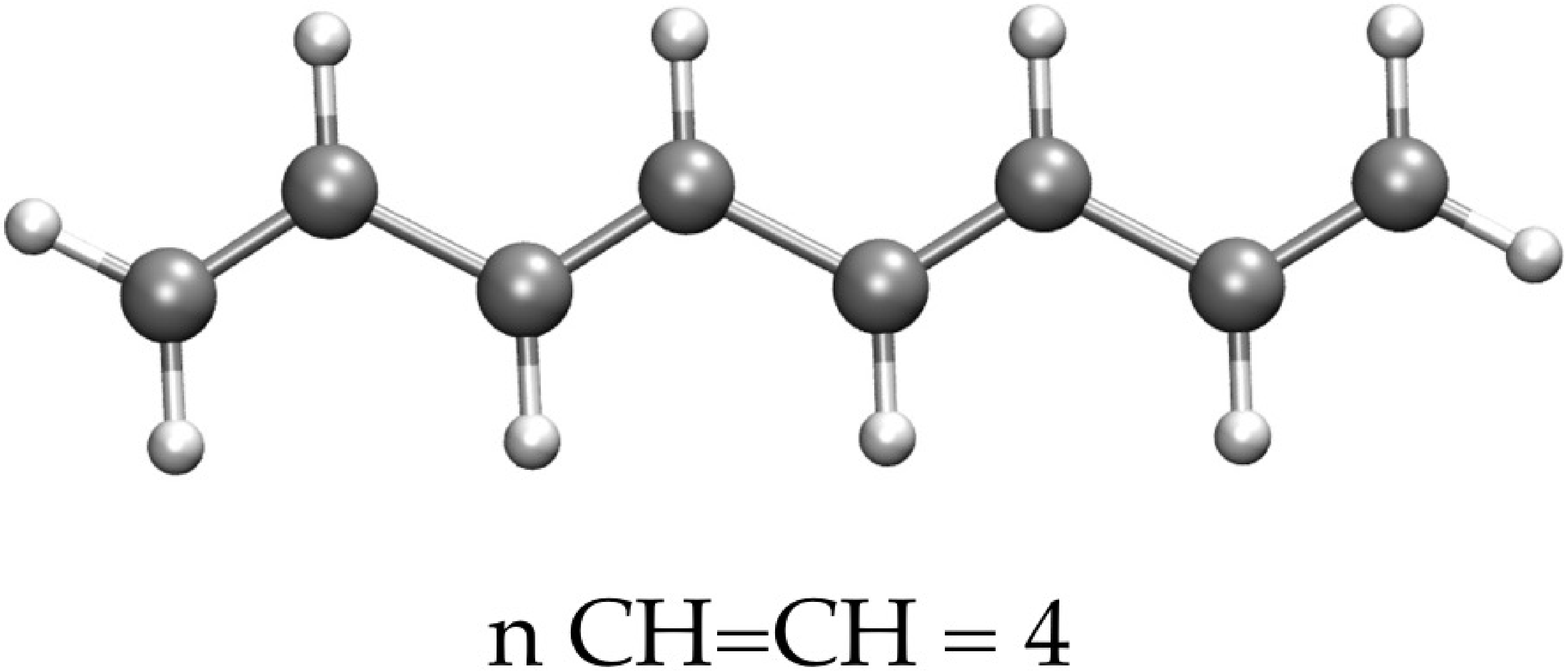}
\end{center}
\end{minipage}
\begin{minipage}[l][7cm][b]{\textwidth}
\resizebox{0.85\textwidth}{!}{\input{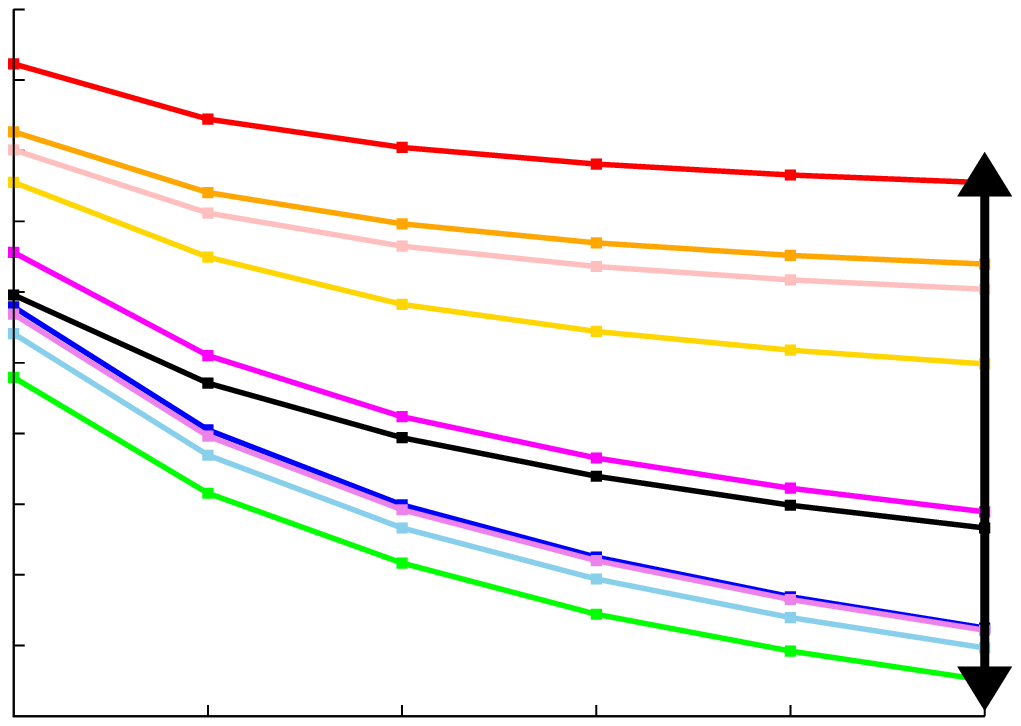}}
\end{minipage}
\caption{Bond Length Alternantion (\AA) in a chain of conjugated double bonds as the length of the chain increases. Calculations carried out with different functionals, Hartree-Fock and CC2.}
\label{BLA}
\end{figure}

Delocalization error is the tendency of approximated methods to over-localize or over-delocalize electron density. The extreme behavior of over-delocalization is given by the Local Density Approximation (LDA). LDA describes the homogeneous electron gas. Hence, it tends to delocalize electrons like in a metal. Semilocal improvements of DFT build on the Local Density Approximation, partially correcting for this feature, but still leading to over-delocalization. On the other extreme, Hartree-Fock (HF) is built to promote electron-pairing, yielding over-localized electrons. It is then easy to see that HF and LDA provide the upper and lower bounds to electron localization. But this electron localization also translates into properties. Let uss see one prototypical example. 
Conjugated double bonds chains are especially prone to this error: whereas HF tends to localize electrons leading to stronger double bonds and weaker single bonds, LDA tends to make all distances similar to each other.
These deviations can be summed up in one single number known as Bond Length Alternation (BLA)\cite{bla}:
\begin{equation}
BLA=\frac{\sum_i l_{db,i}-l_{sb,i}}{i}
\end{equation}
where $l_{db,i}$ and $l_{sb,i}$ are the lengths of adjacent double and single bonds. Large BLAs reveal that double and single bonds are very different in length and vice-versa. The effect of delocalization error can be easily grasped in Figure \ref{BLA}, where the evolution of BLA with the number of CH=CH units is calculated with different methods. 
HF provides the least delocalized and LDA the most delocalized conjugated system. The reference value, CC2, as well as all other methods, fall within the HF/LDA range. The difference between the HF and LDA value, which already at two CH=CH units represents 30\% of the absolute value, is not constant, but increases with the length of the chain, leading to a dramatic difference between the two methods.
In other words, this error becomes crucial for big systems and solid state. 
Moreover, the lack of regularity of this error implies that a systematic scheme for correcting this problem is difficult to implement. 
However, knowing the limiting cases, an estimation of uncertainty can be designed for any crystal.\par

\begin{figure}[H]
\includegraphics[width=0.8\textwidth]{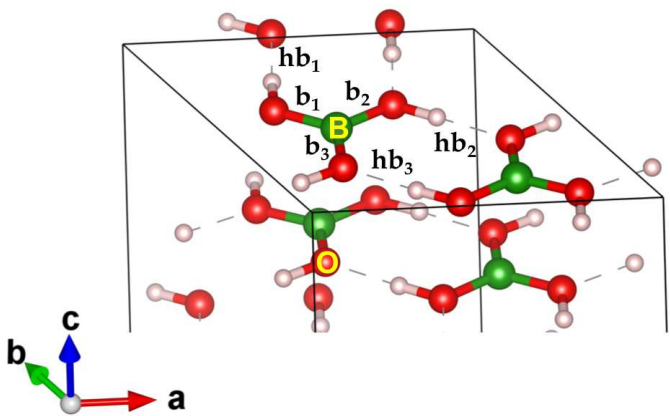}
\caption{Structure of the B(OH)\textsubscript{3} crystal.}
\label{bacid}
\end{figure}

Let us see how delocalization error is transferred to solid state calculations. 
We will consider as an example the crystal structure of boric acid, belonging to space group $P\, 3_2$, whose primitive cell is shown in Figure \ref{bacid}. B(OH)\textsubscript{3} molecules are organized in sheets, parallel to the $a,b$ plane and perpendicular to the $c$ axis. The crystal is stabilized by a strong network of hydrogen bonds, while across-sheet contacts are regulated by electrostatic and dispersion interactions.
 Table \ref{table_bacid} reports the experimental, HF and LDA values of: {\it i)} intramolecular B-O bond distances, {\it ii)} intermolecular O-H hydrogen bonds and {\it iii)} and inter-sheet B-O distance. Results clearly show that intramolecular distances are only slighly affected by the method (b\textsubscript{1} to b\textsubscript{3}). HF over-localization leads to shorter intramolecular bonds than LDA, but the overall accuracy is good and not subject to important deviations. However, the wrong concentration of charge in teh crystal is transmitted to the non-bonded network, so that LDA underestimates non-bonded contacts and HF severely overestimates them. In this case, the difference between both values is as large as 0.4 \AA~   (hb\textsubscript{1} to hb\textsubscript{3}). The difference is even more dramatic when looking at the B-O distance (along $c$) that accounts for across-sheet contacts. In this regard, HF predicts a distance almost 1 \AA~ larger than LDA. 
This huge error in the non-bonded contacts dominates the model errors in a crystal (where non-bonded contacts are ever-present). 
This result is particularly significant because as we will see, it persists when dispersion is included in the calculation and it also applies to a wide range of extended (covalent, ionic) systems, highlighting the relevance of the wrong energy description as a functional of the density and its consequences for computation.

\begin{table}[H]
  \caption{Geometrical parameters of the  B(OH)\textsubscript{3} crystal. Bond labels refer to Figure \ref{bacid}. Distances are in \AA. The structure was resolved at T=297 K.}
  \label{table_bacid}
  \begin{tabular}{lr rr}
    \hline
  Distances& HF & LDA& expt.\cite{boric_acid}\\
    \hline
 b\textsubscript{1}& 1.359& 1.368 & 1.377 \\
 b\textsubscript{2}& 1.358&1.364  & 1.351 \\
 b\textsubscript{3}& 1.357&1.364  & 1.349 \\
 hb\textsubscript{1}& 1.874&1.398&  1.822  \\
 hb\textsubscript{2}& 1.882&1.416&  1.843 \\
 hb\textsubscript{3}& 1.880& 1.411 &  1.911 \\
 B-O& 3.697& 2.758 & 3.187   \\
    \hline
  \end{tabular}
\end{table}

\section{Results and discussion}
\subsection{Ionic solids}

\begin{figure}[H]
\resizebox{\textwidth}{!}{\input{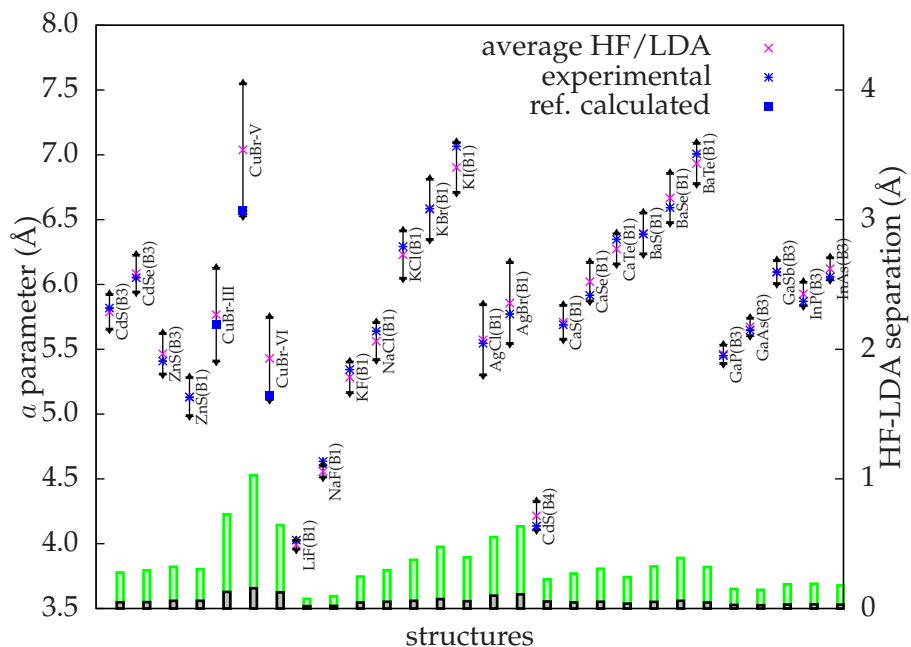}}
\caption{Calculated HF (top) and LDA (bottom) cell parameters for 28 binary ionic solids  with indication of the experimental reference value. Green bars represent the width of the error bar, black bars the same quantity normalized by the experimental cell parameter.}
\label{ionic}
\end{figure}

With the aim of testing the hypothesis that HF and LDA can be used to asses the uncertainty of a calculation, we will start by discussing the properties of a set of ionic solids. The HF and LDA values of cell parameter $a$ are reported in Figure \ref{ionic} (full set of structures and computational data available in ESI). They are invariably the upper and lower bound, respectively, for the experimental data, proving a robust computational error bar.  The green and black bars represent the absolute and normalized amplitude, respectively, of the error bars associated to each structure. Let us have a look at the  LiF-KI family of rocksalt structures. Whereas the absolute error bar in general increases with the size of the cell parameter, the normalized error bar remains fairly constant along the family (with the exception of the smallest structure, LiF). Similar considerations can be drawn for the other families reported in Figure \ref{ionic}, further suggesting that the delocalization error at the origin of the amplidude of the error bar is constant within a given family of structures. This is  extremely interesting because it means that once that the effect of the delocalization error on one member of a given family is known, the corresponding uncertainty for different members of the same family can be quickly estimated (see ESI). To the best of our knowledge, this is the first time that computational errors can be estimated from another compound. Moreover, such estimations are done on the basis of only one compound and can then be applied to other members of the family, meaning that quick \textit{a priori} estimations can be done just based on the symmetry of the structure.

We also tested the HF-LDA bar on less common compounds, such as CuBr. For the III, V and VI phases, HF and LDA parameters have been compared with the accepted reference value, which is itself calculated (Tersoff potential).\cite{CuBr} CuBr III and VI have a B3 and a B1 structure, respectively; and CuBr VI crystallizes in the SC16 tetragonally bonded structure. CuBr shows the highest error bar among the set of ionic compounds in Figure \ref{ionic}, more than 1 \AA. This means that CuBr is extremely sensitive to the computational method, and a much more careful calibration is required than for studying a heavier binary compound like CaTe. Hence, this approach enables to identify those compounds whose computational simulation is more complicated (bigger delocalization error) and whose choice of method should be dealt with care.\par 


\subsection{Molecular solids}
\begin{figure}
\resizebox{\textwidth}{!}{\input{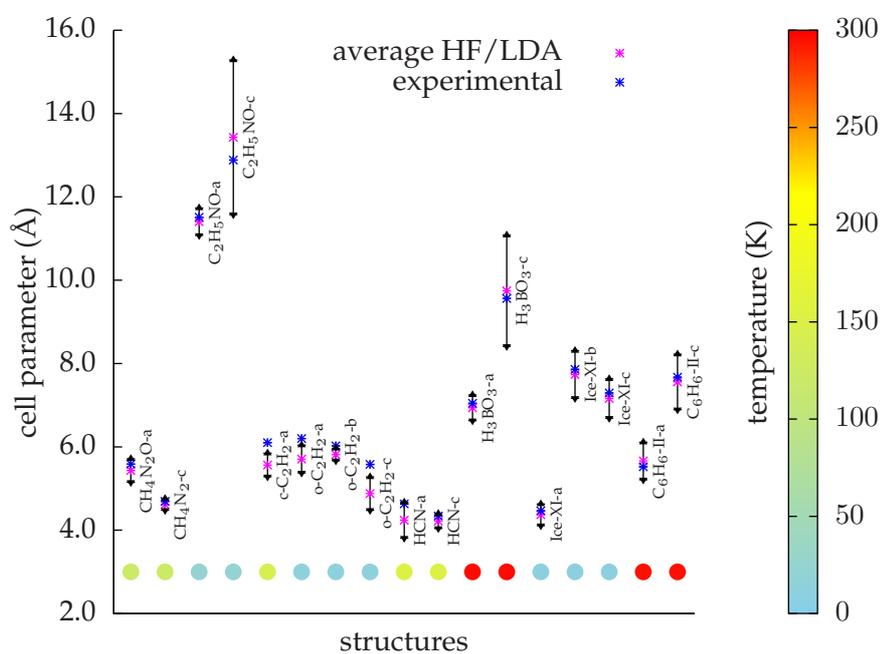}}
\caption{Calculated HF (triangle) and LDA (inverted triangle) cell parameters for 8 molecular solids with indication of the experimental reference value. The temperature corresponding to the experimental observation is reported.}
\label{molecular}
\end{figure}

The discussion of molecular solids is more complex due to several factors, among which thermal expansion and the contribution of non-covalent interactions are particularly relevant. For small molecules, whose intermolecular distances contribute to the cell size (and thus the cell parameters) to a large extent, thermal expansion is significant and can involve volume expansion up to 8\% moving from 0 K to room temperature.\cite{michele16} The standard calculation of cell parameters, which involves a minimization of the potential energy of the crystal, does not account for these thermal effects that have to be included separately, usually by means of the Quasi-Harmonic Approximation (QHA). Additionally, DFT methods are plagued by an insufficient description of dispersion interactions, whose effect can be accounted for with a variety of methods.\cite{disp2011} Dispersion interactions are attractive, and therefore shrink the cell. Overall, these two contributions, thermal expansion and dispersion, go in opposite directions and partially cancel out (see ESI for a full discussion). This is evident when looking at Figure \ref{molecular}, which shows the experimental values and HF-LDA error bars calculated for a series of molecular solids without accounting for thermal or dispersion effects. In spite of the other model errors, all experimental values of cell parameters fall within the error bars, with the only exception of acetylene crystals. 

As shown in Figure \ref{molecular}, the width of the error bar is much larger for molecular solids than for ionic ones. As already discussed for Table \ref{table_bacid}, this is due to the dependence of intermolecular distances on the method employed. Dispersion and thermal effects, which are of extreme importance in the simulation of molecular solids and cannot be neglected, have effects that are some orders of magnitude inferior to the method-dependent error (see ESI). This implies that our simple HF/LDA model, which uses the potential energy and does not account for thermal or dispersion effects, can also be applied to molecular solids for assessing the sensitivity of the system to delocalization error.\par 
\subsection{Transition pressures}

We have shown that the uncertainty associated to DFT calculated cell parameters can be expressed in terms of the HF and LDA values, and that this error is some orders of magnitude higher than the corresponding experimental uncertainty. In this sense, the case of cell parameters is: {\it (i)} a proof-of-concept of the validity of our model, {\it (ii)} an inductive proof of how this difference can be used to estimate how much a given computation depends on the model (functional) chosen.
However, it is not useful in the experiment-theory validation due to the extremely low experimental uncertainty. 

The mutual validation of measured and simulated quantities along with their error bars can be recovered when tackling experimentally less accessible quantities, such as transition pressures. Transition pressures are commonly used to assess the quality of DFT functionals,\cite{Tpressure2006} and they can have an associated experimental uncertainty up to 2.9 GPa even for simple structures.\cite{XA_tp} This uncertainty derives from a complex set of factors, including the accuracy of the  pressure and temperature readings of the sample during the crystallographic measurement.\cite{High_P_cry}\par
\begin{figure}[H]
\resizebox{\textwidth}{!}{\input{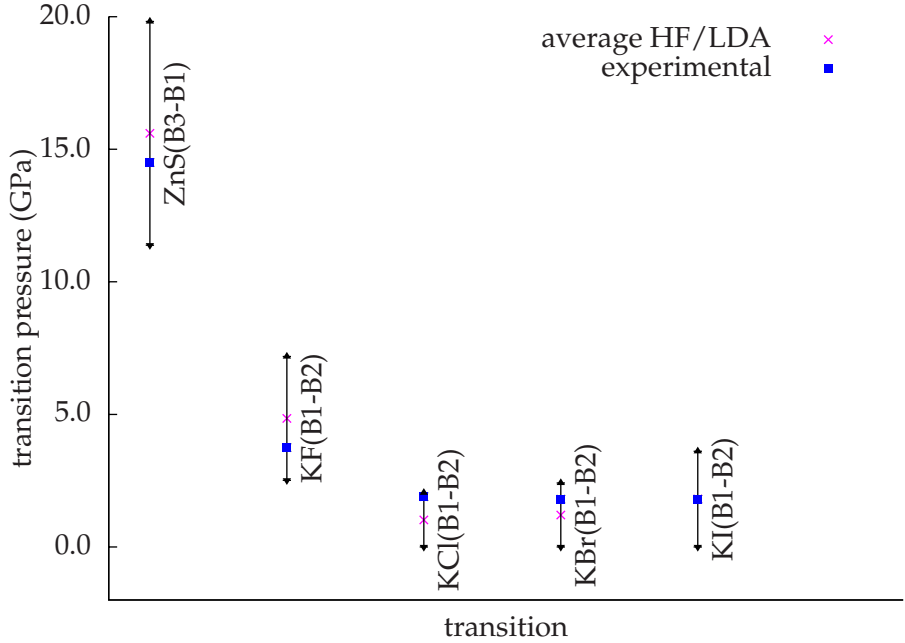}}
\caption{Calculated HF (top) and LDA (bottom) transition pressures (GPa) with indication of the reference experimental value.}
\label{tpres}
\end{figure}

What happens when we look at HF and LDA derived transition pressures? Similarly to what was reported  for cell parameters, LDA underestimates and HF overestimates transition pressures. Figure \ref{tpres} shows the HF and LDA values of a series of transition pressures. For the B1 to B2 transition of alkali halide, we see that again the experimental value falls between the HF and LDA values.\cite{XA_tp} This holds even for large systems (KCl to KI), which are characterized by very low transition pressures. In these cases LDA inverts the order of stability of the two phases, which have been plotted as zero. 
Just like for the cell parameter, it is easy to see that the choice of functional should be handled with much more care for  ZnS than for alkali halides.\citep{ZnS_tp} 
But what is more interesting, in all cases, our computational uncertainty is of the same order of magnitude as the experimental one. Hence, the delocalization error based computational uncertainty can be directly compared with the experimental error bar to mutually asses the quality of the results.

\section{Conclusion}
Summarizing, we showed that the choice of DFT functional for the simulation of solid systems greatly affects both cell parameters and transition pressures. This is attributed to the well-known delocalization error of DFT, and it has been shown that HF and LDA provide a robust error bar for the calculated values. The amplitude of this error bar can be used not only to identify (and even {\it a priori} estimate) which systems are more affected by the density functional used, but in some cases it can also be directly compared with experimental results and their uncertainty. First, we focused on the cell parameters of ionic solids as a proof-of-concept, confirming the validity of our method and showing  that the DFT-derived uncertainty is several orders of magnitude superior to the experimental one. Moving to molecular solids, we showed that also the extent of thermal and dispersion effects affect cell parameters to an extent that is by some orders of magnitude inferior to the delocalization error, which is confirmed as the main source of uncertainty. After validating our model on cell parameters, we focused on transition pressures, and showed that for experimentally less precise data, HF and LDA yield an error bar of the same order of magnitude as the experimental one. In this case, the experimental and calculated transition pressure values can be compared directly along with their associated error bars. The general picture that emerges from this study is that a paradigm shift in the interaction between experimental and computational chemistry is needed. Too often, simulation parameters are tuned case by case to match as closely as possible a given experimental value, when a reasoned approach based on the properties of the solid at hand and its method dependence should be rather preferred. This is because the intrinsic variability associated with the computational method chosen is no lower than the experimental uncertainty, and thus accurately matching the calculated value to the experimental one does not necessarily improve the quality of the simulation. 
Finally, we have also constructed a solid state dataset of experimental cell parameters and transition pressures that can be used for future benchmarking (see ESI). Out of all these data, we have that the computational error bar encloses in most cases the experimental value data we have collected.
Overall, we have devised a simple and robust indicator that provides for any quantity of interest of a material a guide in the choice of the simulation setup.

\bibliographystyle{unsrtnat}
\bibliography{bibliography_solid_state_error}

\begin{thebibliography}{18}
\providecommand{\natexlab}[1]{#1}
\providecommand{\url}[1]{\texttt{#1}}
\expandafter\ifx\csname urlstyle\endcsname\relax
  \providecommand{\doi}[1]{doi: #1}\else
  \providecommand{\doi}{doi: \begingroup \urlstyle{rm}\Url}\fi

\bibitem[Authier(2013)]{early_cryst}
A.~Authier.
\newblock \emph{{Early Days of X-ray Crystallography}}.
\newblock OUP Oxford, 2013.
\newblock ISBN 9780191635014.

\bibitem[Dovesi et~al.(2005)Dovesi, Civalleri, Roetti, Saunders, and
  Orlando]{dov2005}
R.~Dovesi, B.~Civalleri, C.~Roetti, V.~R. Saunders, and R.~Orlando.
\newblock \emph{{Ab Initio Quantum Simulation in Solid State Chemistry}}, pages
  1--125.
\newblock John Wiley \& Sons, Inc., 2005.
\newblock ISBN 9780471720898.

\bibitem[Zobel et~al.(2017)Zobel, Nogueira, and Gonzalez]{patrick_ipea_2017}
J.~P. Zobel, J.~J. Nogueira, and L.~Gonzalez.
\newblock {The IPEA dilemma in CASPT2}.
\newblock \emph{Chem. Sci.}, 8:\penalty0 1482--1499, 2017.

\bibitem[Jeng(2006)]{bias_physics_2005}
M.~Jeng.
\newblock {A Selected History of Expectation Bias in Physics}.
\newblock \emph{Am. J. Phys.}, 74:\penalty0 578--583, 2006.

\bibitem[Herbstein(2000)]{precision_Xray}
F.~H. Herbstein.
\newblock {{How Precise are Measurements of Unit-Cell {\-}Dimensions from
  Single Crystals?}}
\newblock \emph{Acta Crystallogr. A}, 56\penalty0 (4):\penalty0 547--557, 2000.

\bibitem[Bragg and Bragg(1913)]{Bragg13}
W.~H. Bragg and W.~L. Bragg.
\newblock {The Reflection of X-Rays by Crystals}.
\newblock \emph{Proc. Royal Soc. Lond. A Math. Phys. Eng. Sci.}, 88:\penalty0
  428--438, 1913.

\bibitem[{Canc\`es, E} and {Dusson, G.}(2017)]{discretization}
{Canc\`es, E} and {Dusson, G.}
\newblock {Discretization Error Cancellation in Electronic Structure
  Calculation: Toward a Quantitative Study}.
\newblock \emph{ESAIM: M2AN}, 51:\penalty0 1617--1636, 2017.

\bibitem[Cohen et~al.(2008)Cohen, Mori-S{\'a}nchez, and Yang]{yang}
A.~J. Cohen, P.~Mori-S{\'a}nchez, and W.~Yang.
\newblock {Insights into Current Limitations of Density Functional Theory}.
\newblock \emph{Science}, 321:\penalty0 792--794, 2008.

\bibitem[Autschbach and Srebro(2014)]{Srebro2014}
J.~Autschbach and M.~Srebro.
\newblock {Delocalization Error and “Functional Tuning” in Kohn–Sham
  Calculations of Molecular Properties}.
\newblock \emph{Acc. Chem. Res.}, 47:\penalty0 2592--2602, 2014.

\bibitem[Jacquemin and Adamo(2011)]{bla}
D.~Jacquemin and C.~Adamo.
\newblock {Bond Length Alternation of Conjugated Oligomers: Wave Function and
  DFT Benchmarks}.
\newblock \emph{J. Chem. Theory Comput.}, 7:\penalty0 369--376, 2011.

\bibitem[Shuvalov and Burns()]{boric_acid}
Robert~R. Shuvalov and Peter~C. Burns.
\newblock A new polytype of orthoboric acid,
  h\textsubscript{3}bo\textsubscript{3}‐3t.
\newblock \emph{Acta Cryst. C}, 59\penalty0 (6):\penalty0 i47--i49.
\newblock \doi{10.1107/S0108270103009685}.

\bibitem[H{\"o}nerlage(2010)]{CuBr}
B.~H{\"o}nerlage.
\newblock {CuBr: Lattice Constants}.
\newblock In \emph{Landolt-B{\"o}rnstein - Group III Condensed Matter - New
  Data and Updates for III-V, II-VI and I-VII Compounds}, volume 44C. Springer,
  2010.

\bibitem[Cutini et~al.(2016)Cutini, Civalleri, Corno, Orlando, Brandenburg,
  Maschio, and Ugliengo]{michele16}
M.~Cutini, B.~Civalleri, M.~Corno, R.~Orlando, J.~G. Brandenburg, L.~Maschio,
  and P.~Ugliengo.
\newblock {Assessment of Different Quantum Mechanical Methods for the
  Prediction of Structure and Cohesive Energy of Molecular Crystals}.
\newblock \emph{J. Chem. Theory Comput.}, 12:\penalty0 3340--3352, 2016.

\bibitem[Marom et~al.(2011)Marom, Tkatchenko, Rossi, Gobre, Hod, Scheffler, and
  Kronik]{disp2011}
N.~Marom, A.~Tkatchenko, M.~Rossi, V.~V. Gobre, O.~Hod, M.~Scheffler, and
  L.~Kronik.
\newblock {Dispersion Interactions with Density-Functional Theory: Benchmarking
  Semiempirical and Interatomic Pairwise Corrected Density Functionals}.
\newblock \emph{J. Chem. Theory Comput.}, 7:\penalty0 3944--3951, 2011.

\bibitem[Uddin and Scuseria(2006)]{Tpressure2006}
J.~Uddin and G.E. Scuseria.
\newblock Theoretical study of zno phases using a screened hybrid density
  functional.
\newblock \emph{Phys. Rev. B}, 74:\penalty0 245115, 2006.

\bibitem[Potzel and Taubmann(2011)]{XA_tp}
O.~Potzel and G.~Taubmann.
\newblock {The Pressure Induced B1 to B2 Phase Transition of Alkaline Halides
  and Alkaline Earth Chalcogenides. A First Principles Investigation}.
\newblock \emph{J. Solid State Chem.}, 184:\penalty0 1079 -- 1084, 2011.

\bibitem[Hejny and Minkov(2015)]{High_P_cry}
C.~Hejny and V.~S. Minkov.
\newblock {High-Pressure Crystallography of Periodic and Aperiodic Crystals}.
\newblock \emph{IUCrJ}, 2:\penalty0 218--229, 2015.

\bibitem[Chen et~al.(2006)Chen, Li, Cai, and Zhu]{ZnS_tp}
X.~R. Chen, X.~F. Li, L.~C. Cai, and J.~Zhu.
\newblock {Pressure induced phase transition in ZnS}.
\newblock \emph{Solid State Commun.}, 139\penalty0 (5):\penalty0 246 -- 249,
  2006.
\newblock ISSN 0038-1098.

\end{thebibliography}

\end{document}